\documentclass[iop,numberedappendix,twocolappendix,tighten]{emulateapj}

\usepackage[english]{babel}
\usepackage{amsmath}
\usepackage{graphicx}
\usepackage{wasysym}
\usepackage{soul}
\usepackage[dvipsnames]{xcolor}
\usepackage{parskip}
\usepackage{wrapfig}
\usepackage{url}
\usepackage{float}

\begin{document}
\pagestyle{myheadings}
\title[Insolation Around Rapid Rotators]{Gravity-Darkened Seasons: Insolation Around Rapid Rotators}

\author{John P. Ahlers} 
\affil{Physics Department, University of Idaho, Moscow, ID 83844}

\begin{abstract}
I model the effect of rapid stellar rotation on a planet's insolation. Fast-rotating stars have induced pole-to-equator temperature gradients (known as gravity-darkening) of up to several thousand Kelvin that affect the star's luminosity and peak emission wavelength as a function of latitude. When orbiting such a star, a planet's annual insolation can strongly vary depending on its orbit inclination. Specifically, inclined orbits result in temporary exposure to the star's hotter poles. I find that gravity-darkening can drive changes in a planet's equilibrium temperature of up to $\sim15\%$ due to increased irradiance near the stellar poles. This effect can also vary a planet's exposure to UV radiation by up to $\sim80\%$ throughout its orbit as it is exposed to an irradiance spectrum corresponding to different stellar effective temperatures over time. \\
\end{abstract}


\section{Introduction}
A planet's climate is heavily influenced by the type of star it orbits. For example, stellar type determines a planet's exposure to cosmic rays and UV radiation \citep{bruzual1993spectral,griessmeier2009protection}, as well as the system's ice line and habitable zone \citep{traub2011terrestrial}. Planetary atmospheric and climatic behaviors are driven by insolation patterns, which in the right circumstances can result in seasons unlike any in our solar system. This work models insolation around fast-rotating early-type stars and demonstrates potential effects rapid rotation can have on a planet's climate.  

Early-type stars with effective temperatures $\geq6200$ K possess radiative exteriors and almost no magnetic field. As a result, their primordial rotation rates are not magnetically damped \citep{2012ApJ...757...18A}.  Early-type stars therefore often rotate rapidly, which induces pole-to-equator temperature gradients of up to several thousand Kelvin \citep{harrington1968intrinsic,fremat2005effects}. This gradient affects both the star's luminosity and peak emission wavelength as a function of stellar latitude \citep{von1924radiative}.

When orbiting such a star, a planet's seasonal insolation pattern can strongly vary depending on orbit geometry. Specifically, an inclined orbit results in more exposure to the host star's hotter poles, affecting temperature variations over the course of the planet's year. The pole-to-equator stellar flux gradient, called gravity-darkening, can also affect chemical processes in a planet's atmosphere as it is exposed to irradiance corresponding to different stellar effective temperatures over time. This effect could play a major role in the thermal structure, photochemistry, and photoionization of planetary atmospheres \citep{lammer2003atmospheric,ribas2005evolution,yung2005photochemistry}.

Exoplanets orbiting early-type stars frequently misalign from their host star's rotation plane \citep{winn2010hot,2009ApJ...705..683B,ahlers2014spin,2015ApJ...814...67A}. Therefore, gravity-darkened seasons likely occur on a significant number of exoplanets orbiting early-type stars. Understanding this phenomenon is an important step in revealing exoplanet atmospheric and surficial properties in the regime of early-type systems. 

in this paper, I demonstrate how spin-orbit misalignment and gravity-darkening can combine to produce unusual seasonal patterns. In \S2 I derive the insolation model, in \S3 I calculate the insolation of a spin-orbit misaligned planet orbiting a gravity-darkened star and demonstrates its effects on planet equilibrium temperature and received UV flux, and in \S4 I discuss implications for climate and atmospheric behavior. 


\section{Model}
I model gravity-darkened seasons by including the von Zeipel effect \citep{von1924radiative} in my insolation model and test a planet's insolation in various orbit configurations. I use traditional blackbody radiation as the star's emission function because early-type, fast-rotating stars have radiative exteriors and are well-modeled as blackbody radiators \citep{2012ApJ...757...18A}. The total irradiance as a function of wavelength on a planet at any given time is,
\begin{equation}
K(\lambda)=\int_{\phi}\int_{\theta}B(\lambda,T(\theta))
\frac{I(\mu)}{I(1)}\mu R_\odot\sin^2(\theta)d\theta d\phi
\label{eq:irradiance}
\end{equation}

where $B(\lambda,T(\theta))$ is the stellar emission function, $I(\mu)/I(1)$ is the normalized stellar limb-darkening profile, and $\mu$ is a factor to represent the star as a projected disk in the plane of the planet's sky. The integral limits $\phi$ and $\theta$ are traditional azimuthal and polar angles, respectively, with the XY plane defined as the stellar equator. A two-dimensional integral with proper limits of azimuthal angle $\phi$ and polar angle $\theta$ yeilds the instantaneous solar output per wavelength as seen by the orbiting planet. I explain how to handle each element of the above equation in the following subsections and in Appendix \ref{app:deriv} and list static values of the model in Table \ref{table:params}. 

\begin{figure}
\includegraphics[trim=50 210 110 210, clip, width=0.48\textwidth]{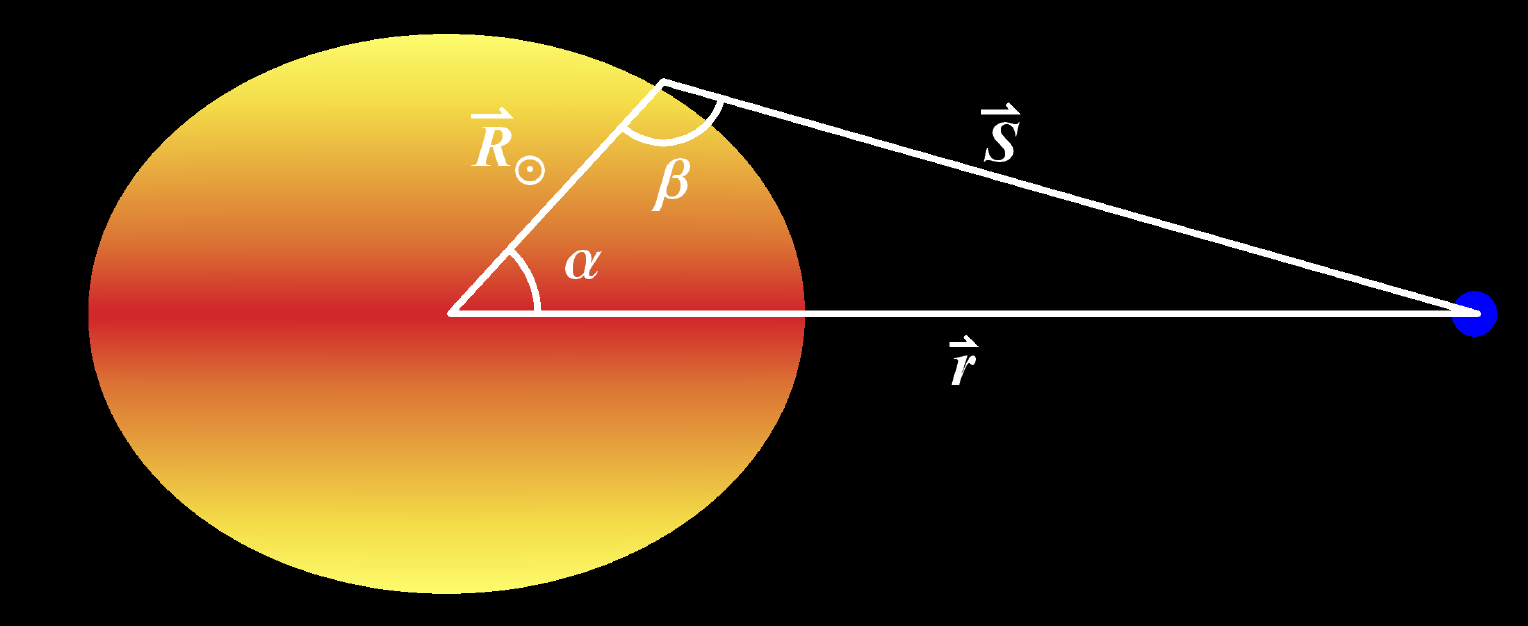}
\caption{\footnotesize Definitions of vectors and angles used in the derivation of Equation \ref{eq:irradiance}. The stellar surface vector $(\vec{R_\odot})$ is not constant in magnitude due to the star's oblateness. The angle $(\pi-\beta)$ describes the angle between the planet's line-of-sight vector $(\vec{S})$ a given location on the stellar surface, which appears in the limb-darkening and rectilinear projection terms in Equation \ref{eq:irradiance}.}
\label{fig:angles}
\end{figure}

\subsection{Stellar Emission}
The stellar emission function $B(\lambda,T(\theta))$ is the function most relevant to the star (e.g. blackbody radiation). The type of emission function in Equation \ref{eq:irradiance} can be interchanged straightforwardly because the star's gravity-darkening effect is handled entirely within the effective temperature function $T(\theta)$. 

A star's rotation induces a pole-to equator gradient in effective surface gravity due to centrifugal force. For fast-rotators, the centrifugal force is enough to significantly lessen the effective gravity near the equator, resulting in oblate stars. This change in surface gravity also produces a temperature gradient across the surface, described by the von Zeipel theorem:
\begin{equation}
T(\theta)=T_{\mathrm{pole}}\left(\frac{g(\theta)}{g_{\mathrm{pole}}}\right)^\epsilon
\label{eq:vonzeipel}
\end{equation}
where $g(\theta)$ is the effective surface gravity as a function of latitude, $g(0)\equiv g_{pole}$ is the surface gravity at the rotation pole, and $\epsilon$ is the so-called gravity-darkening parameter. This parameter is $0.25$ for ideal blackbody radiators and decreases toward zero depending on the radial extent of a star's convective envelope. I derive the stellar temperature function $T(\theta)$ in Appendix \ref{app:temp}.

Stars of spectral type $\sim$F6 or earlier are expected to have radiative exteriors and are well-modeled by blackbody emission; hence, $\epsilon=0.25$ is a reasonable assumption. However, recent observations suggest that $\epsilon$ can deviate signficantly from theory. For example, \citet{Monnier342} measured Altair's gravity darkening parameter at $0.190\pm0.012$. 

\begin{figure}
\centering
\begin{tabular}{r l}
\includegraphics[scale=0.196]{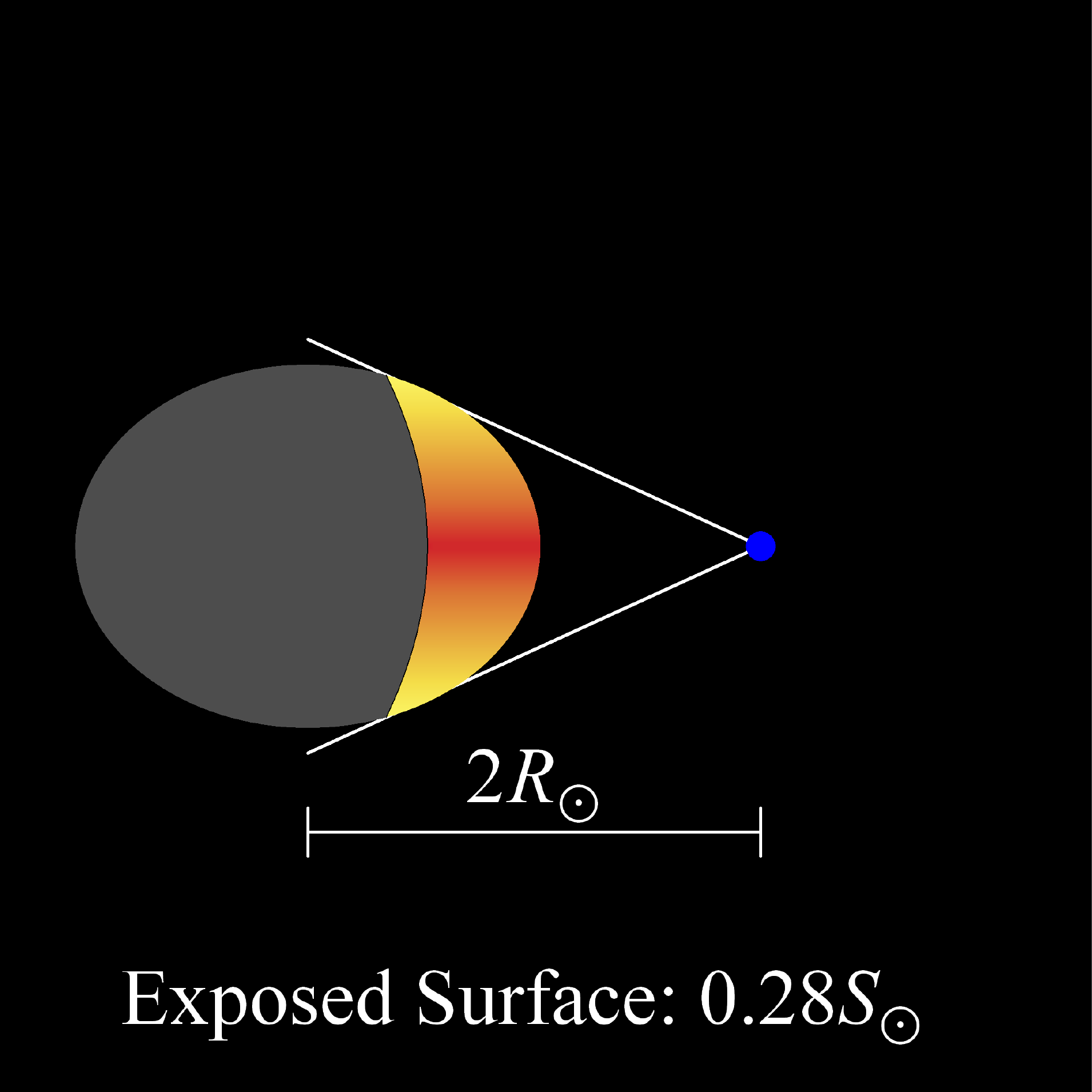} & \includegraphics[scale=0.196]{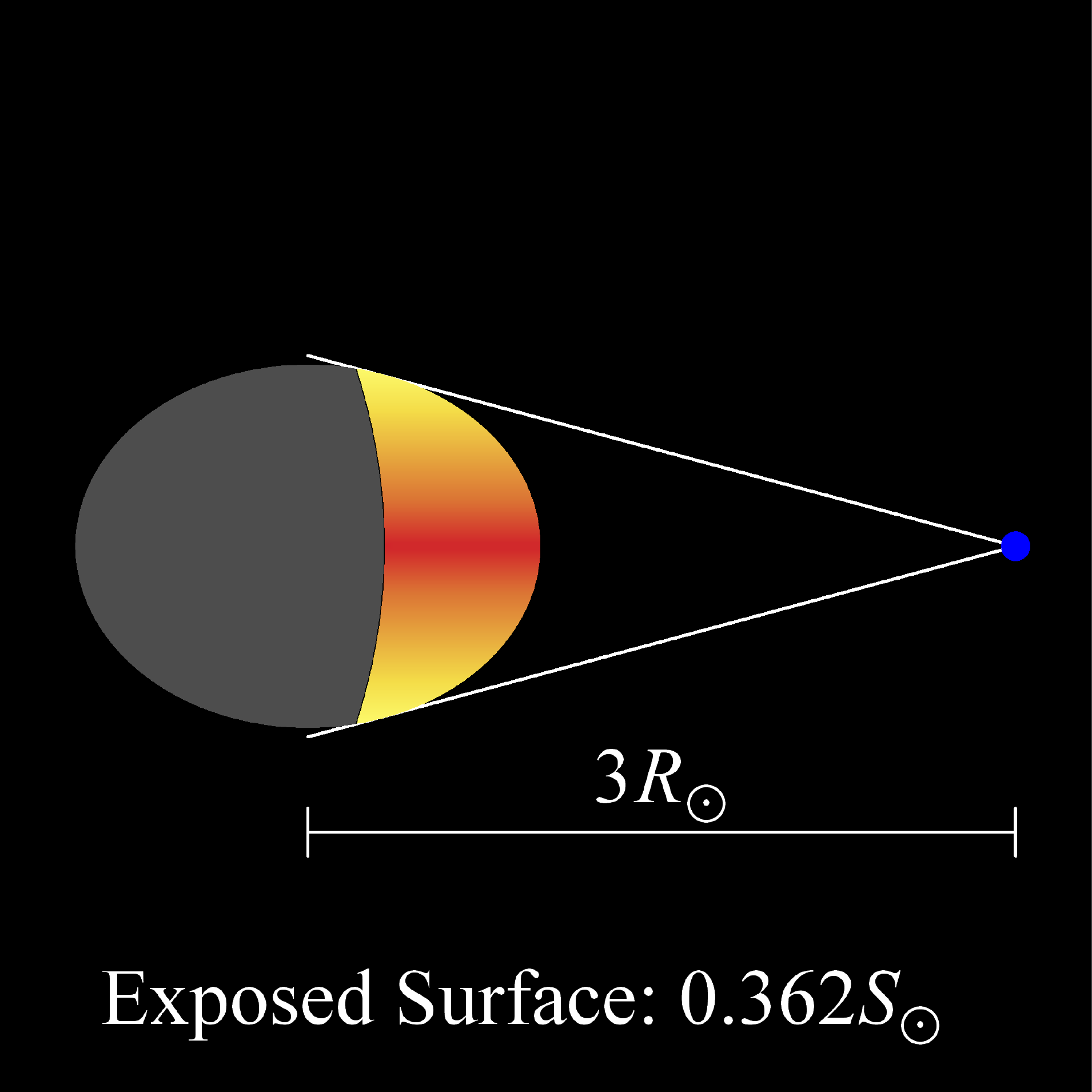} \\
\end{tabular}
\caption{\footnotesize Example of how the total stellar surface area exposed to the planet changes for different orbital distances. The colored-in region of the star represents the area that contributes to the planet's instantaneous irradiance. The border of this area is defined by the region where the line-of-sight vector $\vec{S}$ is tangential to the stellar surface. At $2\mathrm{R_\odot}$, the planet is exposed to $28\%$ of the stellar surface ($S_\odot$), and at $3\mathrm{R_\odot}$ the planet is exposed to $36\%$.} 
\label{fig:exposure}
\end{figure}

\subsection{Limb-Darkening}
Stellar limb-darkening is a brightness effect that stems from the star's optical depth and scale height, which results in the outer limb of a star appearing dimmer than the the center for a given point of view. This effect is well-reproduced, with the emperical formula,
\begin{equation}
\frac{I(\mu)}{I(1)}=1-\sum_{k=1}^\infty a_k(1-\mu^{k/2})
\label{eq:limbdark}
\end{equation}
where $\mu=\cos(\pi-\beta)$ describes the angle between the line of sight and the normal vector of the star's surface (see Figure \ref{fig:angles}). The constants $a_k$ represent limb-darkening coefficients unique to each star; however, several works provide estimates of these coefficients as functions of stellar effective temperature \citep{2010A&A...510A..21S,2011A&A...529A..75C,2013A&A...552A..16C}. 

Typically, linear or quadratic approximations of Equation \ref{eq:limbdark} are employed in stellar models. As long as $\mu$ is known, then any limb-darkening law can be used in Equation \ref{eq:irradiance}. I derive $\mu$ for my chosen coordinate system in Appendix \ref{app:limbdark}.

\subsection{Integral Limits}
Evaluating Equation \ref{eq:irradiance} depends heavily on the correct choice of the integral limits $(\phi,\theta)$. The rotation-induced asymmetry of the star adds two-fold difficulty to a traditional insolation model: the star is no longer spherically symmetric and its effective temperature varies as a function of stellar latitude. Figure \ref{fig:exposure} shows how the planet's location in the system determines what part of stellar surface must be integrated. In general, the limits of integration are set by all $(\phi,\theta)$ that satisfy the inequality,
\begin{equation}
\vec{R_\odot}\cdot\vec{r}\geq R_\odot^2
\label{eq:ineq}
\end{equation}
which I derive in Appendix \ref{app:limits}. This inequality is valid for any position in any orbit configuration, except for the limit of extremely close-in orbits $(r/R_\odot\lesssim 2.2)$, where the planet's size becomes relevant in determining insolation by latitude.


\section{Results}\label{sec:results}
\begin{figure}
\includegraphics[trim={0 0 0 0.5cm},scale=0.7]{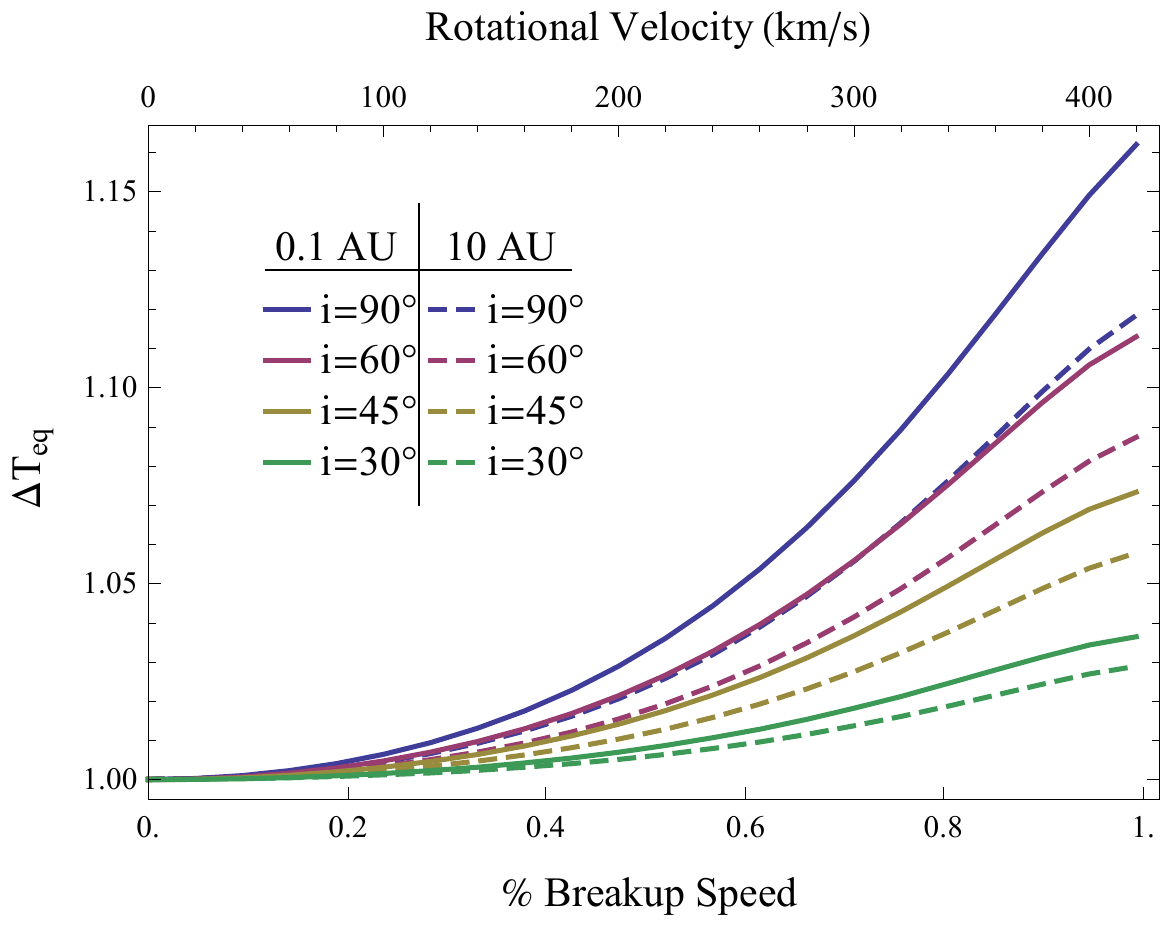}
\caption{\footnotesize Fractional change in the planet's equilibrium temperature versus stellar rotation rate throughout the course of the planet's orbit for various inclinations and orbit distances. The fractional change in temperature corresponds to the planet being primarily exposed to the hotter poles or cooler equator due to its orbit geometry. In general, close-in, highly inclined planets experience the strongest induced temperature changes, but even modestly inclined planets in the outer solar system can undergo significant gravity-darkened seasons. Early-type stars frequently rotate near their breakup speed; for example, Vega and Regulus both rotate at near $90\%$ of their breakup speeds \citep{0004-637X-708-1-71,0004-637X-628-1-439}.}
\label{fig:spinup}
\end{figure}

\begin{figure*}
\centering
\begin{tabular}{r l}
\includegraphics[scale=0.683]{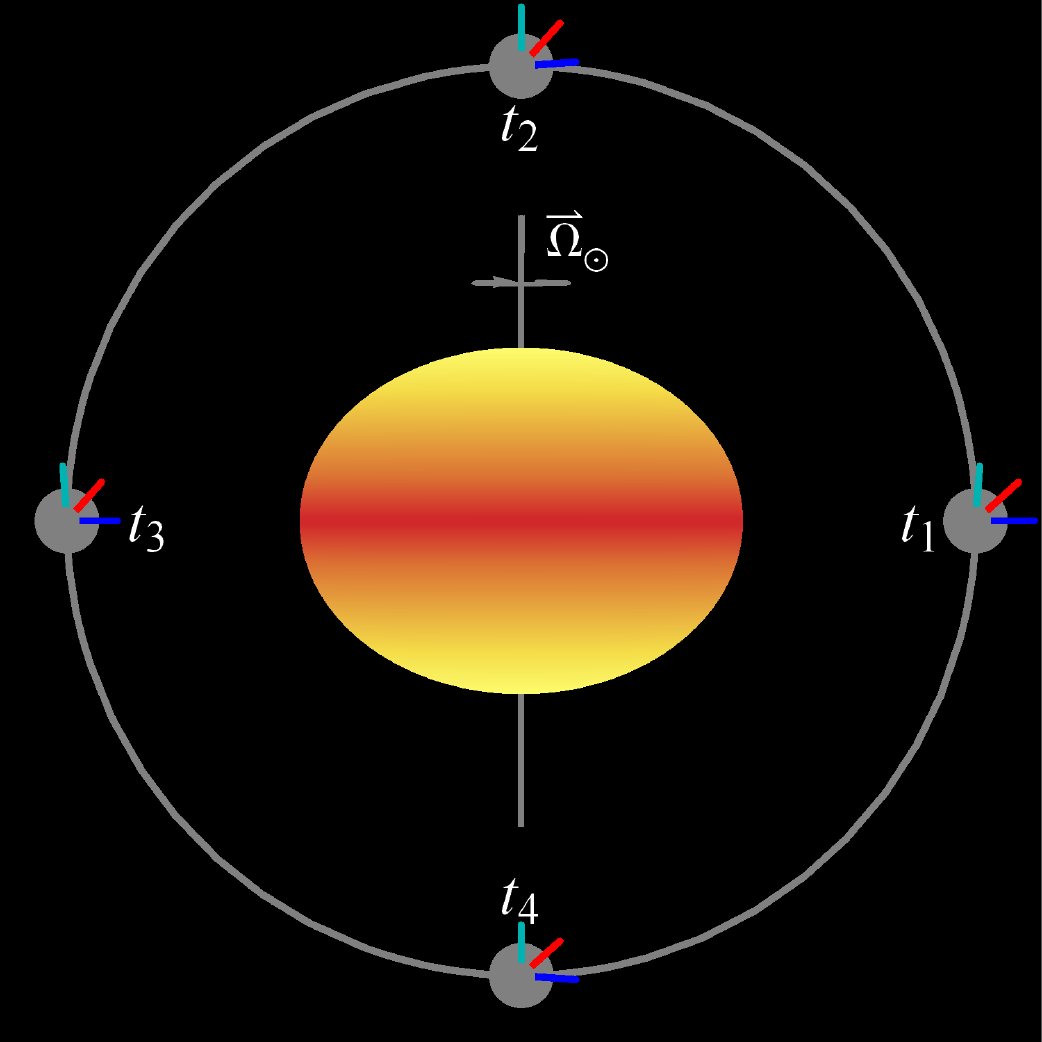} & \includegraphics[scale=0.78]{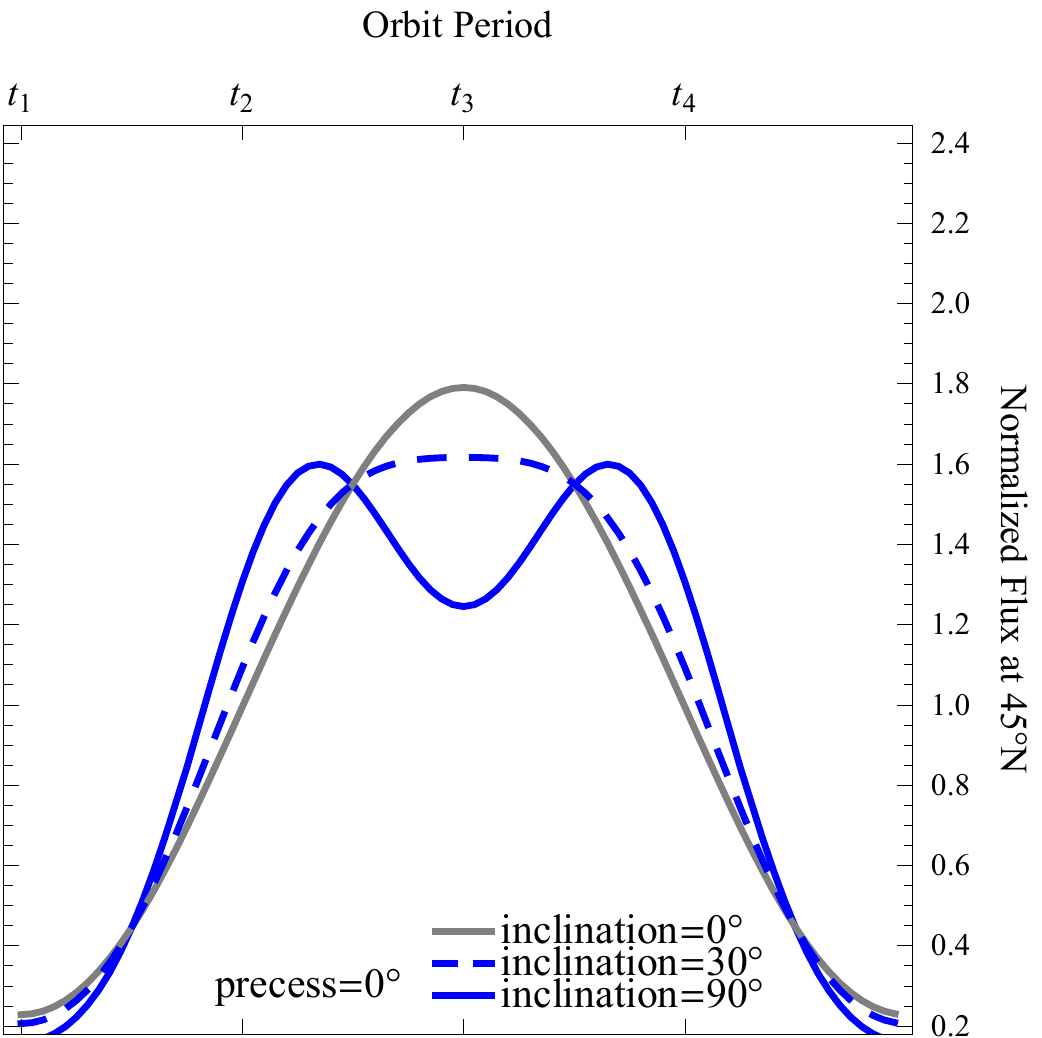} \\
\includegraphics[scale=0.78]{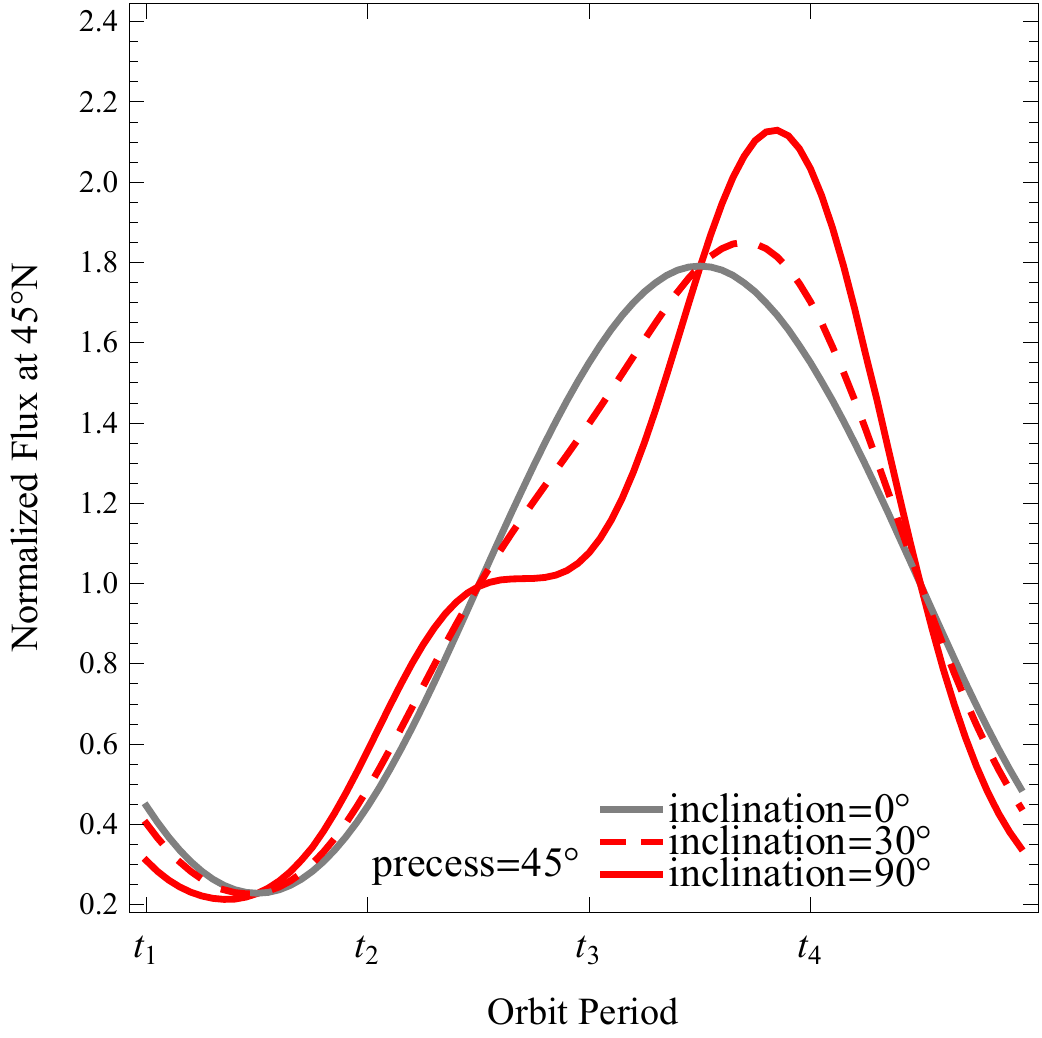} & \includegraphics[scale=0.78]{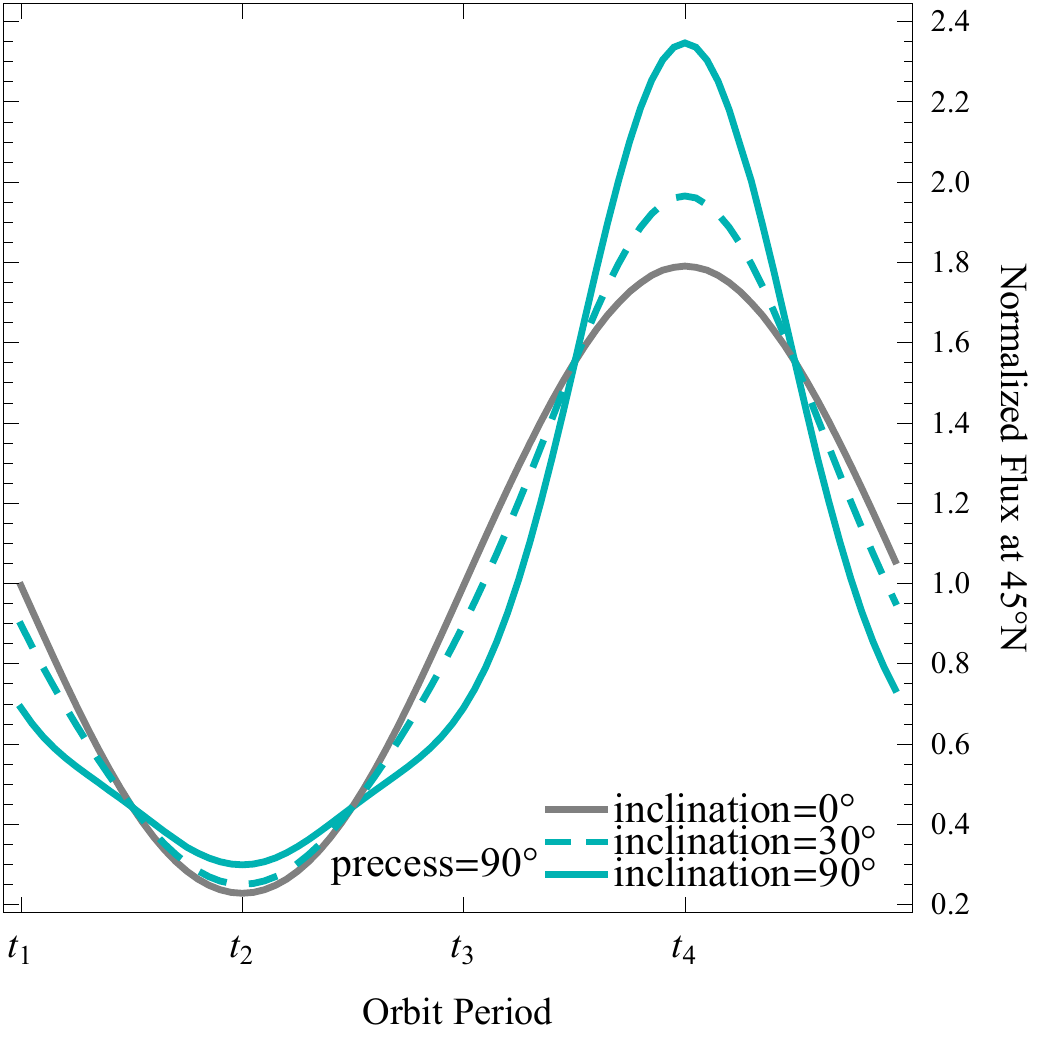} \\
\end{tabular}
\caption{\footnotesize Insolation at $45^\circ$ North latitude throughout an orbit for different precession angles. The blue, red, and cyan plots respectively correspond to precession angles of $0^\circ$, $45^\circ$, and $90^\circ$ relative to the planet's longitude of ascending node. All three configurations have obliquities of $30^\circ$ and include gravity-darkening induced changes in flux for orbit inclinations of $0^\circ$, $30^\circ$, and $90^\circ$. For all configurations, variations away from tradiational insolation patterns scale with inclination and stellar rotation rate (see Figure \ref{fig:spinup}). The different precession angles produce significantly different seasonal patterns due to combining with the gravity-darkening effect at different phases. A precession angle of $90^\circ$ results in mild winters and extreme summers, while a precession angle of $0^\circ$ produces mild summers and extreme winters and, at high inclinations, two distinct hottest times of the year. Figure \ref{fig:obliquity} demonstrates the effect of multiple insolation peaks at different planet obliquities.}
\label{fig:latitude}
\end{figure*}

\begin{figure}
\includegraphics[trim=0 10 0 -7, clip, width=0.453\textwidth]{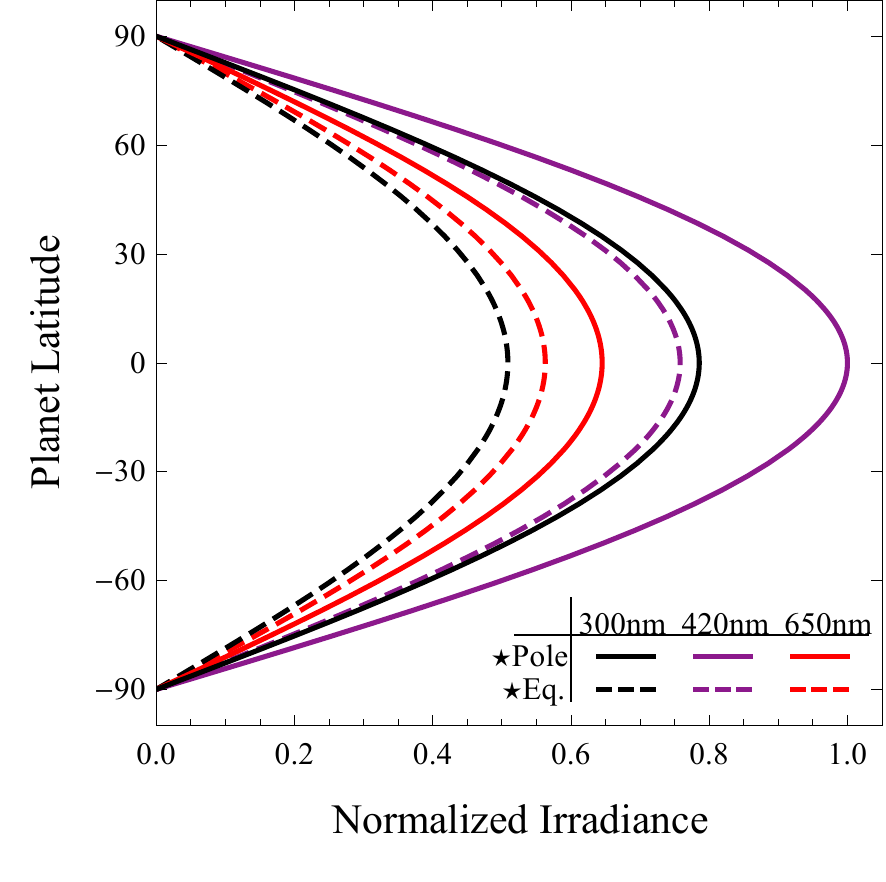}
\caption{\footnotesize Normalized irradiance across the surface of a planet undergoing gravity-darkened seasons. The solid and dashed lines show irradiance by planet latitude when closest to the stellar pole and stellar equator, respectively. The incoming solar flux is less at all wavelengths when near the equator. The most drastic change in flux is in UV wavelengths, where intensity can change by as much as 80\% throughout its orbit. These changes in UV irradiance occur at twice the orbit frequency.}
\label{fig:irradiance}
\end{figure}

I apply the gravity-darkened insolation model to a synthetic system using the parameters listed in Table \ref{table:params}. I demonstrate how gravity-darkened seasons are affected by stellar rotation rate in Figure \ref{fig:spinup}. I demonstrate possible seasonal patterns for various orbit configurations in Figure \ref{fig:latitude} and show how the planet's irradiance by wavelength can change in Figure \ref{fig:irradiance}.
 
I find two insolation characteristics unique from planets orbiting solar-type stars. First, nonstandard patterns in the planet's total received solar flux and equilibrium temperature occur throughout its year; the nature of these patterns depends on the planet's inclination and direction of axial tilt, or precession angle. Second, the insolation's spectral energy distribution varies over time due to being exposed to emission corresponding to the hotter stellar poles or cooler equator. 

\renewcommand*{\arraystretch}{1.2}
\begin{table}
\centering
\begin{tabular}{l l} \hline \hline
{\bf Stellar Parameters} & {\bf Value} \\ \hline
$M_\odot$ &  $2.0~\mathrm{M_\star}$\\
$R_\mathrm{eq}$ & $2.1~\mathrm{R_\star}$ \\
$T_\mathrm{pole}$ & 7700 K \\
$a_1$ & 0.19 \\
$a_2$ & 0.36 \\ 
$\zeta$ & 0.23 \\
$\epsilon$ & 0.25 \\ \hline
{\bf Planet Parameters} & {\bf Value} \\ \hline
$a$ & $0.5\mathrm{AU}$ \\
$e$ & $0$ \\
$i$ & $90^\circ$ \\
$\omega$ & $0^\circ$ \\
$\Omega$ & $0^\circ$\\
$A$ & 0 
\end{tabular}
\caption{\footnotesize Static parameters (unless otherwise indicated) used when calculating the planet's insolation in Figures \ref{fig:spinup}, \ref{fig:latitude}, and \ref{fig:irradiance}. Appendix \ref{app:definitions} lists definitions of all variables. The planet's semimajor axis and inclination and the star's rotation rate are listed with each simulation. The stellar limb-darkening coefficients $a_1$ and $a_2$ follow \citet{2010A&A...510A..21S}.  }
\label{table:params}
\end{table}

Using a blackbody emission function and quadratic limb-darkening, I find that an inclined planet's equilibrium temperature can vary by as much as $\sim15\%$ throughout the course of its orbit. This would, for example, correspond to variations in equilibrium temperate between $\sim300\mathrm{K}-345\mathrm{K}$ on a planet near the habitable zone.  Figure \ref{fig:spinup} models how stellar rotation rate drives planetary temperature change for inclined orbits. This change in temperature is caused purely by gravity-darkening and stellar oblateness; effects such as planet albedo or orbital eccentricity were not considered in this study. 

\begin{figure*}
\centering
\begin{tabular}{r l}
\includegraphics[scale=0.78]{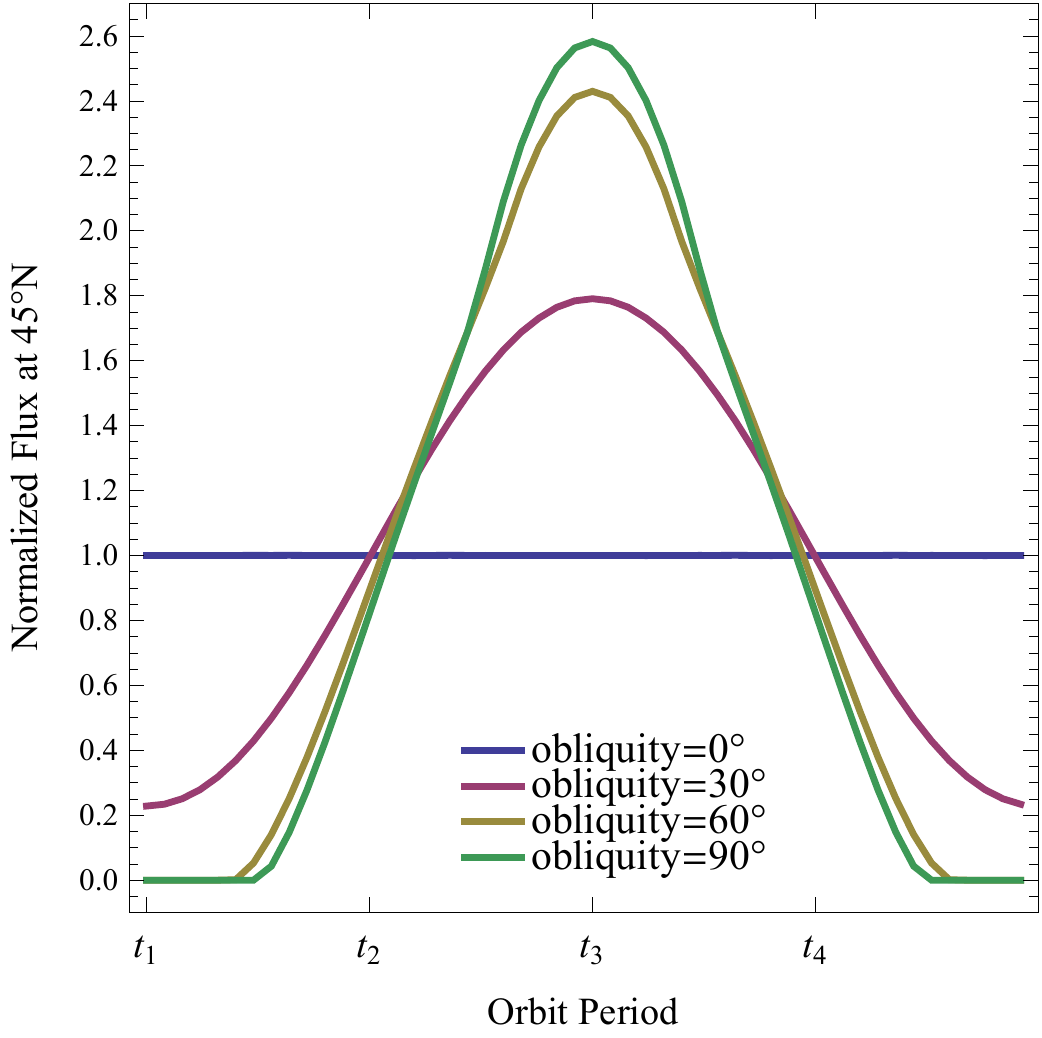} & \includegraphics[scale=0.78]{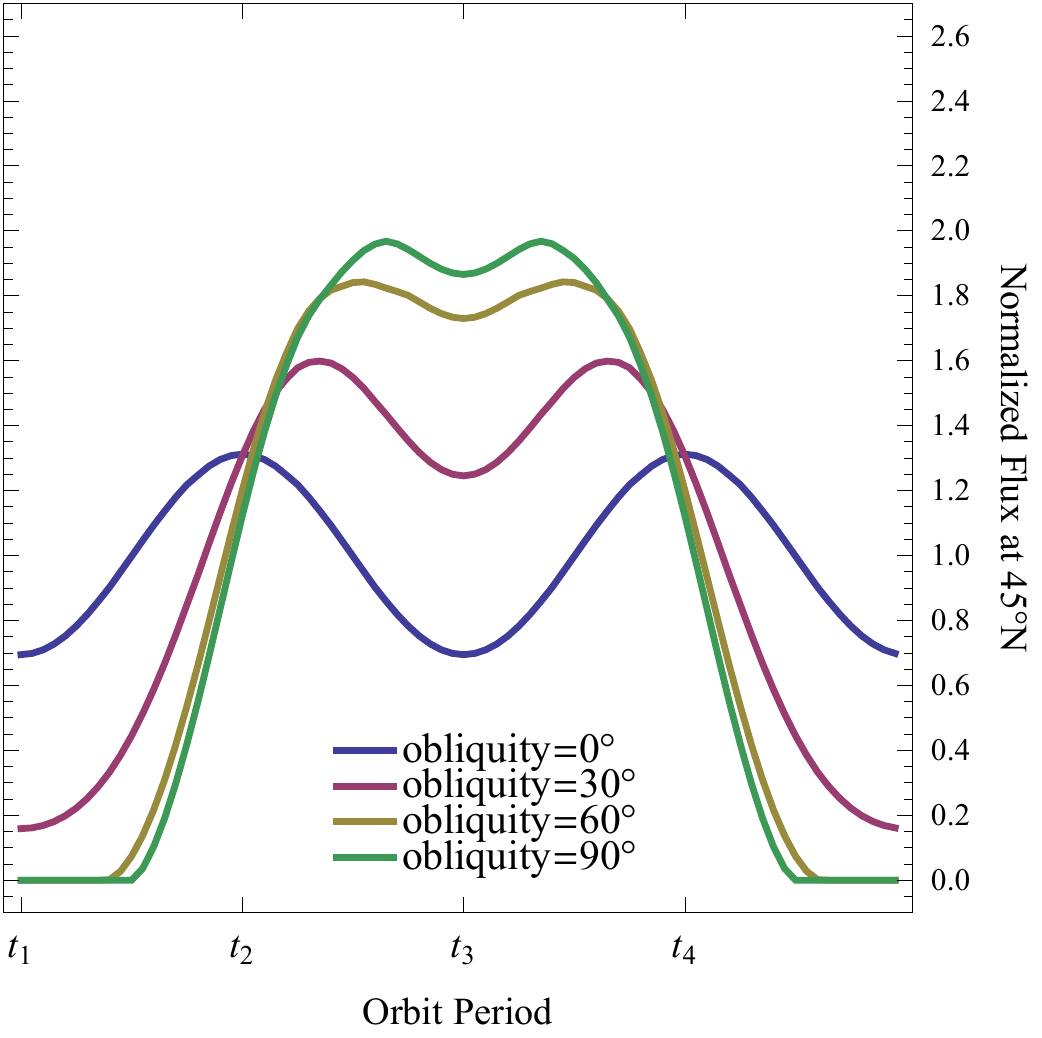} \\
\end{tabular}
\caption{\footnotesize Yearly normalized flux at $45^\circ$ North latitude discluding (left) and including (right) gravity darkening effects. Both plots include yearly insolation values with respect to $0^\circ$, $30^\circ$, $60^\circ$, and $90^\circ$ planet obliquity values. For both sets of integrations, I set the planet's precession angle at $\rho=0^\circ$ and inclination at $i=90^\circ$ (see Figure \ref{fig:latitude}). The left plot shows traditional insolation patterns around a spherically symmetric star. The right plot demonstrates that gravity-darkened seasons occur at all obliquity values. At low obliquities, irradiance varies as a sinusoid, effectively producing traditional seasons at twice the orbit frequency.}
\label{fig:obliquity}
\end{figure*}

The gravity-darkening effect can combine with traditional seasonal effects brought about by a nonzero planet obliquity, resulting in abnormal seasons. Traditionally, a planet's obliquity causes more/less light exposure for a given latitude throughout its orbit. However, inclined orbits around gravity-darkened stars receive more total irradiance each time they pass over one of the stellar poles -- twice per orbit. Gravity-darkening produces planetary temperature changes at twice the frequency of the planet's traditional seasons. These two effects combined can result in unusual seasonal behaviors (Figure \ref{fig:latitude}). I compare gravity-darkened seasons with obliquity-driven seasons in Figure \ref{fig:obliquity} and derive the relevant calculations in Appendix \ref{sec:obliquity}.

The rotation-induced temperature gradient across the stellar surface results in the planet receiving different emission intensities throughout its orbit. This shift is especially evident in the ultraviolet for early-type stars. Figure \ref{fig:irradiance} displays an inclined planet's normalized wavelength-dependent insolation when exposed to the stellar equator and stellar pole.


\section{Discussion \& Conclusion}\label{sec:discussion}

\subsection{Climate Effects}
The equilibrium temperature of an inclined planet around a gravity-darkened star can vary by as much as $\sim 15\%$ throughout its year due to changing total solar irradiance. This effect is additive with traditional seasons -- hemispherical temperature changes brought about by a planet's obliquity. Traditional seasons occur once per orbit, but gravity-darkened seasons occur twice per orbit -- how these two effects coincide plays a large role in determining the planet's seasonal behaviors. 

Ultimately, the nature of gravity-darkened seasons is driven by the phase difference between the planet's precession angle and longitude of ascending node. If traditional summer/winter occurs near the stellar poles, the planet experiences hot summers and mild winters. If traditional summer/winter instead occur near the stellar equator, mild summers and extreme winters occur, with unusually warm spring/autumn seasons. In fact, Figure \ref{fig:spinup} shows that the gravity darkening effect can overpower seasonal temperature changes caused by obliquity such that traditional spring and autumn are hotter than a hemisphere's summer, producing two distinct peak heating seasons. 

This temporal heterogeneity in total solar irradiance would likely drive radiative forcing on an Earth-like planet, directly impacting its sea surface temperature and hydrological cycle. For example, as the climate warms, its atmosphere would hold more water vapor, increasing greenhouse gases and further increasing the planet's temperature \citep{held2000water,forster2007changes}. The reverse would hold true when the climate cooled. Changes in total irradiance could also affect giant planet deflation/inflation rates \citep{podsiadlowski1993planet,fortney2011discovery}.  This starkly constrasts with insolation in our solar system, where total solar irradiance varies by only $\sim0.2\%$ over 11-year cycles \citep{haigh2007sun}.  

The equilibrium temperature changes due to gravity darkening shown in Figure \ref{fig:spinup} are maximum values -- in reality, this effect would be mitigated by the planet's albedo, thermal inertia, and atmosphere. The planet would likely not be able to circulate heat globally as quickly as its total irradiance changed, especially for close-in planets.  For example, 55 Cancri e is an exoplanet with observed poor global heat transport \citep{demory2016map}. However, the general trends in Figure \ref{fig:spinup} would still be driven by the planet's changing exposure to sunlight intensity. 

Figures \ref{fig:latitude} and \ref{fig:obliquity} demonstrate how a planet's precession angles and obliquities can affect seasonal insolation patterns when orbiting a gravity-darkened star. These values can change throughout a planet's lifetime. For example, Earth's rotation axis precesses every 26,000 years and oscillates in magnitude every 41,000 years \citep{lissauer2012obliquity,barnes2016obliquity}. A spin-orbit misaligned planet undergoing these changes in axial tilt would be driven through the different insolation scenarios in Figure \ref{fig:latitude} on its precession timescale. Obliquity variations could drive Milankovich cycles whose nature depends on orbit geometry. Future studies of these phenomena could help reveal planetary processes driven by gravity-darkened seasons for the first time.

Recent works on habitable planet Proxima Centauri b \citep{2016Natur.536..437A} offer a path for characterizing exoplanets in detail. By constraining the planet's formation and migration history, high-energy irradiance, incoming stellar particle winds, and tidal interactions, along with the host star's evolution history, one can estimate the planet's atmospheric loss rate, its water budget, and its overall climate regime \citep{ribas2016habitability,turbet2016habitability}. \citet{barnes2016habitability} and \citet{meadows2016habitability} demonstrate that a planet's geologic behavior can be constrained by modeling its orbit evolution and tidal history, as well as heavy element abundances in the planets core. Such works provide possible next steps toward characterizing the nature of exoplanets in early-type systems.   

\subsection{Atmospheric Effects}
Figure \ref{fig:irradiance} shows how the irradiance by wavelength on a spin-orbit misaligned exoplanet orbiting a gravity-darkened star can vary throughout its orbit. These changes occur at all wavelengths, with the strongest variations occuring at wavelengths lower than the peak emission wavelength (near UV-violet for early-type stars). The total UV irradiance can vary by as much as 80\% throughout an exoplanet's year, with the changes occuring near-sinusouidally at twice the orbit frequency.    

Variations in a planet's UV irradiance play a significant role in its photochemistry \citep{forster2007changes}. UV light drives the production of ozone in the Earth's stratosphere \citep{caldwell1994stratospheric}. UV irradiation also plays a significant role in the atmosphere of Saturn's moon Titan, driving much of the organic chemistry in its atmosphere and producing large amounts of aerosols \citep{szopa2006pampre}. Extreme UV irradiation can drive loss processes in an exoplanet's atmosphere. Hydrogen-rich exoplanets under extreme ultraviolet radiation may evaporate down to their cores \citep{lammer2003atmospheric}. 

\citet{forster2007changes} shows how even very small changes in UV irradiation on the Earth can have significant impacts on the structure of its atmosphere. Gravity-darkening can cause massive changes in UV irradiance throughout an inclined planet's orbit; future photochemical and radiative tranfer models could reveal the full impact of gravity-darkening on a planet's atmopshere.   

\subsection{Conclusion}
With rapid stellar rotation and planet spin-orbit misalignment common in early-type systems  \citep{winn2010hot,2012ApJ...757...18A}, gravity-darkened seasons likely occur in a significant number of exoplanets. I quantify how this phenomenon scales with stellar rotation rate, planet inclination, and semi-major axis and shows that a planet's equilibrium temperature can nominally vary by as much as 15\%. 

Such a planet's total solar influx varies at twice its orbit frequency. This work shows how traditional seasons caused by planet obliquity can combine with its changing irradiance, and demonstrates how planet obliquity and gravity-darkening can combine to produce unusual seasonal patterns. In early-type systems, these effects are strongest in UV irradiance, which can have profound impacts on a planet's atmosphere.  

The insolation patterns modeled in this work represent a preliminary investigation into the nature of planets orbiting fast-rotating stars. As planet detection and characterization techniques improve, more and more planets undergoing gravity-darkened seasons will likely be revealed. Future atmospheric models could reveal how gravity-darkened seasons can affect a planet's climate and photochemistry, shedding new light on planets orbiting stars dissimilar to our own.



\bibliographystyle{apj}
\bibliography{citations}


~\\
~\\
~\\
~\\
~\\
\appendix

\section{Derivations} \label{app:deriv}

\subsection{Effective Temperature Function}\label{app:temp}

To second order, the stellar effective surface gravity is the gradient of the total surface potential,
\begin{equation}
\vec{g}=-\vec{\nabla}\left[\frac{-GM_\odot}{R_\odot}\left(1-\frac{J_2R_{eq}^2P_2(\mu)}{R_\odot}  \right) -\frac{1}{2}\Omega_\odot^2R_\odot^2\sin^2(\theta)  \right]
\label{eq:gravity}
\end{equation}

where $R_{eq}$ is the star's equatorial radius, $J_2$ is the second-order gravitational harmonic term dictated by the star's oblateness, $P_2(\mu)$ is a second-order Legendre polynomial, and $\Omega$ is the star's rotation rate. This gradient produces a two-component vector: 
\begin{equation}
\vec{g}=g_r\hat{r}+g_\theta\hat{\theta}
\end{equation}

Converting these terms to cartesian coordinates,
\begin{equation}
\begin{array}{l}
g_x=g_r\cos(\phi)\sin(\theta)-g_\theta\sin(\theta) \\
g_y=g_r\sin(\phi)\sin(\theta)+g_\theta\cos(\theta) \\
g_z=g_r\cos(\theta)
\end{array}
\end{equation}

The total effective gravity does not depend on the azimuthal angle $\phi$, so $g\equiv g(\theta)=\sqrt{g_x^2+g_y^2+g_z^2}$. With $g(\theta)$, the star's effective temperature distribution is known through Equation \ref{eq:vonzeipel}. The expression for temperature can then be inserted into any stellar emission function.

\subsection{Limb-Darkening and Rectilinear Projection}\label{app:limbdark}

Figure \ref{fig:angles} shows the angle $\beta$ between the planet's line of sight and a given location on the stellar surface. The law of cosines gives $\mu=\cos(\pi-\beta)$ as
\begin{equation}
\mu=\frac{r^2-R_\odot^2-S^2}{2R_\odot S}
\label{eq:mu}
\end{equation}

The planet's orbit vector is:
\begin{equation}
\vec{r}=r\left(\begin{array}{c}
\cos(\Omega)\cos(\omega+f)-\sin(\Omega)\sin(\omega+f)\cos(i) \\
\sin(\Omega)\cos(\omega+f)+\sin(\Omega)\sin(\omega+f)\cos(i) \\
\sin(\omega+f)\sin(i) \\
\end{array} \right)
\label{eq:r}
\end{equation}

Modelling the star as an oblate spheriod gives the stellar radius:
\begin{equation}
\vec{R_\odot}=\frac{Req}{\sqrt{\sin^2(\theta)+\frac{cos^2(\theta)}{(1-\zeta)^2}}}
\left(\begin{array}{c}
\cos(\phi)\sin(\theta)\\
\sin(\phi)\sin(\theta)\\
\cos(\phi)\\
\end{array} \right)
\label{eq:R}
\end{equation}

where $\zeta$ is the star's oblateness and $R_{eq}$ is the star's equatorial radius. The stellar oblateness can be derived via the Darwin-Radau relation \citep[e.g.,][]{bourda2004precession}. $S$ can be expressed in terms of $r$ and $R_\odot$ via,
\begin{equation}
S=r^2+R_\odot^2-2rR_\odot\cos(\alpha)
\end{equation}
where 
\begin{equation}
\cos(\alpha)=\frac{\vec{r}\cdot\vec{R_\odot}}{|\vec{r}||\vec{R_\odot}|}
\end{equation}
Backsolving, the limb-darkening angle $\mu$ can be expressed in terms of the stars polar and azimuthal angles $(\phi,\theta)$ and the planet's orbital elements. This same factor $\mu$ appears again in Equation \ref{eq:irradiance} outside of the limb-darkening term. This extra factor projects the stellar area of the parameters as a rectilinear disk in the plane of the planet's sky, and is necessary to properly represent the stellar projected area exposed to the planet at any given moment. 

\subsection{Integral Limits}\label{app:limits}
The limits of $(\phi,\theta)$ are determined by the line-of-sight vector $\vec{S}$. From the planet's point of view, the stellar edge of visibility is set according to where $\vec{S}$ is tangential to the stellar surface $(\vec{S}\cdot\vec{R}=0)$. The angle $(\pi-\beta)$ is constrained to $-\pi/2\leq\pi-\beta\leq\pi/2$. With $|\pi-\beta|\geq\pi/2$, the following inequality
\begin{equation}
r^2\geq R_\odot^2+S^2
\label{eq:limitineq}
\end{equation}
is true for the region of the star exposed to the planet. Using the law of cosines,
\begin{equation}
S^2=r^2+R_\odot^2-2(\vec{r}\cdot\vec{R_\odot})
\label{eq:cos}
\end{equation}
Inputting Equation \ref{eq:cos} into Equation \ref{eq:limitineq}, a useable inequality describing the limits of $(\phi,\theta)$ is obtained (Equation \ref{eq:ineq}). Equations \ref{eq:r} and \ref{eq:R} can be employed to evaluate this inequality. Depending on the type of numerical integrator being used, this inequality can be applied to Equation \ref{eq:irradiance} as a boolean statement or, more elegantly, by inserting dynamic functions $(\phi(\theta),\theta(\phi))$ as limits of integration.

\subsection{Planet Equilibrium Temperature}
Traditionally, a planet's equilibrium temperature is straightforward to calculate. However, the gravity-darkening effect can cause a planet to be exposed to different stellar effective temperatures and stellar projected areas throughout the course of its orbit -- therefore, the stellar luminosity (as seen by the planet) can change over time. An approximate value of the instantaneous ``effective" stellar luminosity can be expressed as an integral of Equation \ref{eq:irradiance} over all wavelengths of the exposed part of the star,
\begin{equation}
\widehat{L}_\odot=\int_0^\infty K(\lambda) d\lambda
\end{equation}
The integral limits given by Equation \ref{eq:ineq} apply to this integral. The total effective luminosity as seen by the planet is then,
\begin{equation}
L_\odot=\widehat{L}_\odot\frac{S}{\widehat{S}}
\label{eq:luminosity}
\end{equation} 
where $S$ is the total stellar surface area of the oblate spheroid and $\widehat{S}$ is the stellar surface area exposed to the planet. The effective luminosity can then be used to approximate the planet's equilibrium temperature for a given part of its orbit,
\begin{equation}
T_{eq}\simeq\left( \frac{L_\odot (1-A)}{16\sigma\pi r^2} \right)^{1/4}
\end{equation}

where $A$ is the planet's albedo and $\sigma$ is the Stefan-Boltzmann constant. 

\subsection{Insolation and Planet Obliquity} \label{sec:obliquity}
Figures \ref{fig:latitude} and \ref{fig:obliquity} show normalized irradiance at $45^\circ$ North latitude for different orbit geometries and axial tilts. I account for planet obliquity $(\eta)$ and precession angle $(\rho)$ by adopting the derivation from \citet{mcgehee2012paleoclimate}. I start with a point $u$ on the surface of the planet in spherical coordinates,
\begin{equation}
u=\left(\begin{array}{c}
\cos(\varphi)\cos(\gamma)\\
\cos(\varphi)\sin(\gamma)\\
\sin(\varphi)
\end{array} \right)
\label{eq:planetcoords}
\end{equation}
where $\varphi$ is planet latitude and $\gamma$ is planet longitude. I rotate this point on the surface by planet obliquity $\eta$ and planet precession angle $\rho$,
\begin{equation}
\hat{u}=\left(\begin{array}{c c c}
\cos(\rho) & -\sin(\rho) & 0\\
\sin(\rho) & \sin(\rho) & 0\\
0 & 0 & 1\\
\end{array} \right) \left(\begin{array}{c c c}
\cos(\eta) & 0 & \sin(\eta)\\
0 & 1 & 0\\
-\sin(\eta) & 0 & \cos(\eta)
\end{array} \right) u
\label{eq:obliquity}
\end{equation}

which gives the incident angle of any point on the planet's surface. I combine the effective luminosity seen by the planet at any time with its obliquity to find the incoming flux $F(\varphi,\gamma)$ of the planet as a function of planet latitude and longitude via,
\begin{equation}
F(\varphi,\gamma)=\int\frac{L_\odot(\hat{r}\cdot\hat{u})}{4\pi r^2}\mathrm{d}\gamma
\label{eq:flux}
\end{equation}
where $L_\odot$ is given by Equation \ref{eq:luminosity} and $\hat{r}$ is the unit vector of the planet's orbit, given by Equation \ref{eq:r}.

To generate Figure \ref{fig:latitude}, I set the planet latitute to $45^\circ$ North and integrate equation \ref{eq:flux} with respect to all longitudes receiving irradiation. I perform this calculation at different points in the planet's orbit to find its normalized flux. Planet rotation rate does not affect this calculation because the integration includes all substellar longitudes.

\section{Parameter Definitions}\label{app:definitions}

\begin{tabular}{l l} \hline \hline
{\bf Parameter} & {\bf Definition} \\ \hline
$A$ & Planet's albedo \\
$a$ & Planet's semi-major axis \\
$B(\lambda,T(\theta))$ & Stellar emission function \\
$e$ & Planet's eccentricity \\
$f$ & Planet's true anomaly \\
$I(\mu)$ & Stellar limb-darkening profile  \\ 
$i$ & Planet inclination (spin-orbit misalignment) \\
$J_2$ & Gravitational harmonic constant \\
$M_\odot$ & Stellar mass  \\ 
$R_\odot$ & Stellar radius (Eq. \ref{eq:R}) \\
$R_{eq}$ & Star's equatorial radius   \\
$u$ & Planet coordinates (Eq. \ref{eq:planetcoords}) \\
$\vec{r}$ & Planet's orbit radius (Eq. \ref{eq:r}) \\ 
$\vec{S}$ & Line-of-sight vector (Fig. \ref{fig:angles}) \\
$S_{\odot}$ & The star's surface area \\
$\alpha$ & angle between $\vec{r}$ and $\vec{R_\odot}$ (Fig. \ref{fig:angles})  \\ 
$\beta$ & angle between $\vec{S}$ and $\vec{R_\odot}$ (Fig. \ref{fig:angles})  \\  
$\gamma$ & Planet's longitude (Eq. \ref{eq:planetcoords})\\
$\epsilon$ & Gravity-darkening parameter (Eq. \ref{eq:vonzeipel})\\
$\zeta$ & Star's oblateness \\
$\eta$ & Planet's obliquity (Eq. \ref{eq:obliquity}) \\
$\theta$ & Star's polar angle  \\ 
$\lambda$ & Stellar emission wavelength \\
$\mu$ & Rectilinear projection factor (Eq. \ref{eq:mu}) \\ 
$\rho$ & Planet's precession angle (Eq. \ref{eq:obliquity}) \\
$\sigma$ & Stefan-Boltmann constant \\
$\phi$ & Star's azimuthal angle  \\
$\varphi$ & Planet's latitude (Eq. \ref{eq:planetcoords})\\ 
$\Omega$ & Planet's longitude of ascending node \\
$\Omega_\odot$ & Star's angular rotation rate \\
$\omega$ & Planet's argument of pericentre
\end{tabular}

\end{document}